\immediate\write18{makeindex \jobname.nlo -s nomencl.ist -o \jobname.nls}

\documentclass[journal,10pt]{IEEEtran}
\usepackage{amssymb}
\usepackage{adjustbox}
\usepackage{hyperref}
\usepackage{cite}
\usepackage{graphicx}
\usepackage{amsmath}
\usepackage{algorithm}
\usepackage{algorithmic}
\usepackage{array}
\usepackage{fixltx2e}
\usepackage{stfloats}
\usepackage{nomencl}
\makenomenclature

\begin{document}

\title{Dimensioning and Power Management of Hybrid Energy Storage Systems for Electric Vehicles with Multiple Optimization Criteria}

\author{Huilong Yu,~\IEEEmembership{Member,~IEEE,}
    Francesco Castelli-Dezza,~\IEEEmembership{Member,~IEEE,}
	Federico Cheli,
	Xiaolin Tang$^*$, ~\IEEEmembership{Member,~IEEE,}
	Xiaosong Hu$^*$, ~\IEEEmembership{Senior Member,~IEEE,} and 
	Xianke Lin,~\IEEEmembership{Member,~IEEE}
	
	\thanks{This work was in part supported by National Natural Science Foundation of China (Grant No. 51875054 and No. 51705044) and Chongqing Natural Science Foundation for Distinguished Young Scholars (Grant No. cstc2019jcyjjq0010), Chongqing Science and Technology Bureau, China ( Corresponding authors: X. Hu and X. Tang).}
	\thanks{H. Yu was with the Department of Mechanical Engineering, Politecnico di Milano, 20156, Milano, Italy. He is now with University of Waterloo,N2L 3G1, Waterloo, Canada (e-mail: huilong.yu@uwaterloo.ca).}
	\thanks{F. Cheli and F. Castelli Dezza are with the Department of Mechanical Engineering, Politecnico di Milano, 20156, Milano, Italy (e-mail:federico.cheli@polimi.it;  francesco.castellidezza@polimi.it).}
	\thanks{X. Hu and X. Tang are with the State Key Laboratory of Mechanical Transmissions and the Department of Automotive Engineering, Chongqing University, Chongqing 400044,	China (e-mail: xiaosonghu@ieee.org; tangxl0923@cqu.edu.cn).}
	\thanks{X. Lin is with the Departement of Automotive, Mechanical and Manufacturing Engineering at the Ontario Tech University, Oshawa, ON L1G 0C5, Canada (email:xianke.lin@uoit.ca).}
}

{}
\maketitle

\begin{abstract}
Hybrid energy storage systems that combine lithium-ion batteries and supercapacitors are considered as an attractive solution to overcome the drawbacks of battery-only energy storage systems, such as high cost, low power density, and short cycle life, which hinder the popularity of electric vehicles. A properly sized hybrid energy storage system and an implementable real-time power management system are of great importance to achieve satisfactory driving mileage and battery cycle life. However, dimensioning and power management problems are quite complicated and challenging in practice. To address these challenges, this work proposes a Bi-level multi-objective design and control framework with the non-dominated sorting genetic algorithm-II and fuzzy logic control as key components, to obtain an optimal sized hybrid energy storage system and the corresponding optimal real-time power management system based on fuzzy logic control simultaneously. In particular, a vectorized fuzzy inference system is devised, which allows large-scale fuzzy logic controllers to run in parallel, thereby improving optimization efficiency. Pareto optimal solutions of different hybrid energy storage systems incorporating both optimal design and control parameters are obtained and compared to show the achieved enhancements of the proposed approach.
\end{abstract}


\IEEEpeerreviewmaketitle

\mbox{}
\nomenclature{$V_{bat}$}{Battery voltage ($ V $)}
\nomenclature{$E_0$}{Battery voltage constant ($ V $)}
\nomenclature{$K $}{Battery polarization resistance ($ \Omega $) }
\nomenclature{$Q_{\max } $}{Battery  total capacity ($Ah$)}
\nomenclature{$i$}{Battery current ($A$)}
\nomenclature{$R_{bat} $}{Battery  internal resistance ($ \Omega $) }
\nomenclature{$A $}{ Battery voltage amplitude ($ V $)}
\nomenclature{$B $}{Time constant inverse of the exponential zone ($Ah^{-1}$)}
\nomenclature{$P_{reqbat} $}{ Net requested power from the battery ($kW$)}
\nomenclature{$ N_{bat}$}{Number of  battery cells }
\nomenclature{$\eta_{AD}$}{Efficiency of the DC/AC converter}
\nomenclature{$m_{\textit {HESS}}$}{Total mass of the HESS ($kg$)}
\nomenclature{$A_{cl}$}{Pre-exponential factor of battery cycle life model}
\nomenclature{$R_{cl}$}{Gas constant}
\nomenclature{$T  $}{The absolute temperature }
\nomenclature{$A_h $}{Ah-throughput of the battery }
\nomenclature{$V_{ct} $}{Total open circuit voltage of the supercapacitor pack ($ V $)}
\nomenclature{$R_{sct} $}{Total equivalent series resistance of the supercapacitor pack ($ \Omega $) }
\nomenclature{$R_{sc}$}{Series resistance of one supercapacitor ($ \Omega $) }
\nomenclature{$P_{reqsc}$}{Demand power from the supercapacitor pack ($kW$)}
\nomenclature{$\eta_{dc}$}{Efficiency of the DC/DC converter }
\nomenclature{$C_{sct}$}{ Total capacity of the supercapacitor pack}
\nomenclature{$x_{SOE}$}{State of energy }
\nomenclature{$V_{ctmax}$}{The initial open circuit voltage ($ V $)}
\nomenclature{$V_c$}{Open circuit voltage of one supercapacitor ($ V $)}
\nomenclature{$N_{sc}$}{Total number of the banks}
\nomenclature{$P_{sc}$}{Actual total output power of one supercapacitor ($kW$)}
\printnomenclature

\section{Introduction}\label{sec:intro}
\IEEEPARstart{C}{hallenges} of air pollution, fossil oil crisis, and greenhouse gas emissions have attracted unprecedented attention from governments, academia, and industries around the world on electric vehicles (EVs). After the rapid development over the past decade, the worldwide promotion and application of EVs have reached a considerable scale. However, the dynamic performance, cost, and durability of EVs are still closely related to the design, integration, and control of the energy storage systems (ESSs) \cite{Hu2015}. The high cost and short cycle life of battery-only ESSs have become one of the biggest obstacles to the further penetration of the EVs     \cite{Hu2019aa}.Lithium-ion battery-only ESSs with high energy density and relatively good power density have become the dominant choice for powering EVs. However, battery degradation can be accelerated when high discharging/charging power demands are required during operation \cite{suri2016control,Hu2018aaa}. In contrast, supercapacitors (SCs) can tolerate much more charging/discharging cycles and exhibit superior ability to cope with high power demands due to their special energy storage mechanisms. However, their low energy density hinders their large-scale application in EVs  \cite{burke2000ultracapacitors,ZHANG2018b}. A hybrid energy storage system (HESS) that combines both lithium-ion batteries and supercapacitors is considered as one of the most promising solutions to solve the above-mentioned problems in the battery-only or SC-only energy systems \cite{Lukic2006,cao2012new,Ma2015a}. The configuration of a HESS varies with different connections of the battery, supercapacitor and DC/DC converter. The HESS in this work is the most studied configuration that uses a bi-directional DC/DC converter to connect the supercapacitor with the battery in parallel. In this case, the voltage of the supercapacitor can be adjusted over a wider range  \cite{cao2012new}. Existing research has demonstrated that HESSs can dramatically improve the braking energy recuperation efficiency, eliminate the need for battery over-sizing, and reduce the weight and cost of the entire system \cite{Hochgraf2014}. However, the application of HESS involves complicated sizing and power management  problems \cite{yu2016real}. The following three paragraphs will review this topic in terms of HESS sizing, power management, and combined sizing and power management.

Finding the optimal number of supercapacitor banks and battery cells that can minimize the cost, mass, energy consumption or battery degradation of the HESS is the so-called HESS sizing problem. A sample-based global  Dividing RECTangles (DIRECT) optimization algorithm is implemented to solve a  multi-objective sizing problem of the HESS \cite{shen2014}, and a rule-based power-split control strategy is implemented to evaluate all design solutions. The non-dominated sorting genetic algorithm II (NSGA-II) is applied to obtain the Pareto front of battery degradation and total cost in designing a semi-active HESS \cite{Song2014a}, where the power split is regulated directly by a devised hardware topology. \cite{Hu2015a, Hu2014c} proposed the use of a convex optimization algorithm to solve the sizing problem of different HESSs, where the size and power control strategy are optimized simultaneously off-line. The Pareto front of energy storage size and fuel economy based on the Bandwidth control strategy is obtained using a parallel-mode multi-objective genetic algorithm in  \cite{Shahverdi2016}, where the controller is also predesigned. NSGA-II is implemented in \cite{Zhang2018} to find the Pareto front of cost, weight, and state of health of a HESS with a predesigned wavelet-transform based power management algorithm as the control law.

Energy management strategies (EMSs) for HESSs aim to distribute energy demand among different energy sources to achieve the desired performance. Both rule-based and optimization-based approaches are comprehensively studied in the previous work  \cite{TIE201382}. A real-time utility function-based control for a semi-active HESS was proposed in  \cite{Yin2015} by formulating a weighted multi-objective optimization problem. Then the formulated problem  is solved based on the Karush-Kuhn-Tucker (KKT) conditions. \cite{Khaligh2014} proposed an  power management strategy for a HESS based on the fuzzy logic supervisory wavelet-transform frequency decoupling approach, which aims to maintain the state of energy (SOE) of the supercapacitor at the optimal value  to increase the power density of the ESS  and prolong the battery lifetime. An explicit model predictive control system for a HESS was proposed and implemented in \cite{Hredzak2015} to make HESS operate within  specific constraints while distributing  current changes with different ranges and frequencies between the supercapacitor and battery. \cite{Zhang2017} developed a real-time predictive power management control strategy based on neural networks and particle swarm optimization algorithm to minimize the integral cost including battery degradation and energy consumption. A variable charging/discharging threshold method  and an adaptive intelligence technique based on historical data  were proposed in \cite{Jia2017} to improve the power management efficiency and smooth the load of a HESS. Two real-time power management strategies based on KKT conditions and neural network were investigated and validated by the experiment work  in \cite{shen2016} to improve the battery state of health performance of a HESS effectively.  A real-time genetic algorithm based power management strategy is proposed in \cite{Wieczorek2017} to optimize the energy efficiency of a HESS. The impact of the proposed strategy on energy consumption, battery current, and cycle cost were analyzed. \cite{Veneri2018} designed and compared rule-based EMSs for a HESS. An optimization-based strategy called $\lambda$-control was devised to maximize the energy efficiency of HESSs in \cite{Castaings2016, Capasso2018}. The $\lambda$-control was implemented and validated in real-time tests. \cite{Wu2020} proposed an artificial potential field-based power allocation strategy with a compensator for battery/supercapacitor HESS.

Most of the aforementioned work was focused on either sizing or power management, in which case, only the sub-optimal solution was obtained due to the reduced searching space. To this end, some researchers combined these two problems and tried to optimize the sizing parameters and control law parameters simultaneously. The combined sizing and power management problem was formulated as a nonlinear programming problem and solved by an open source nonlinear programming solver IPOPT in \cite{DeCastro2012b} and \cite{Araujo2014}. A similar approach can also be found in \cite{Hu2015a}, where the sizing and power management problem was formulated as an integrated convex optimization problem and solved by CVX, which is a Matlab software for disciplined convex programming. Reference \cite{Song2018} formulated and solved the sizing and power management problem of a HESS off-line based on the Pontryagin's maximum principle. The above reviewed combined approaches are off-line, which are quite useful as the reference when design EMSs but not appropriate for real-time implementation. Only a few studies investigated the combined sizing and real-time power management optimization problem. However, the developed real-time algorithms were either too simple or only single-objective. Despite the combined approach, the optimality of the solution in  \cite{Eldeeb2019} was still limited due to the simple design of the power splitting method. DIRECT algorithm was used to find the degree of hybridization and the parameters of a fuzzy logic-based controller simultaneously  \cite{Li2009}. It’s a pity that this work considered only energy efficiency as the objective and the designed membership functions and fuzzy rules were relatively simple. 

In summary, the sizing and control of a HESS should consider multiple objectives, including its lifespan, energy efficiency and capacity. Combined sizing and power management helps maximize the searching space for better optimality. Besides, parametric real-time EMS design is essential for further optimization and practical application. Fuzzy logic control (FLC)  is real-time, adaptive and  intelligent \cite{Cabrera2015, Sabahi2016, Wang2015}. It allows  different operators to merge nonlinearities and uncertainties in the best way and incorporate heuristic control in the form of if-then rules. Its effectiveness and distinct advantages in power management of HESS have been demonstrated, especially after optimizing its membership functions \cite{Yu2017, Li2009}. However, the optimization of these membership functions usually causes low computation efficiency, especially when the designed FLC is complicated. 

Based on our previous studies \cite{Yu2017, Yu2018a}, the primary goal of this work is to develop an efficient multi-objective optimal sizing and real-time power management algorithm for a HESS. The contributions that distinct this work from the previous efforts are: 1) A systematic Bi-level multi-objective optimal sizing and control framework incorporating the battery dynamic model, supercapacitor model,  evaluation model and adaptive parametric real-time EMS is proposed; 2) A parametric vectorized fuzzy inference system is devised for the first time to the best of our knowledge, which allows a large number of fuzzy logic controllers running in parallel and enables fast training and optimization; 3) Pareto optimal solutions of a HESS for an electric race car are obtained using NSGA-II under the proposed framework. The corresponding optimal sizing parameters and membership functions of a real-time fuzzy logic-based EMS are obtained simultaneously.

Sizing and control of the HESS for an electric race car is investigated as a case study in this work, which aims to minimize the operating cost for a racing team and to reduce the environmental impact   caused by the waste battery. The remainder of this work is divided into six parts. Section 2  elaborates on the proposed Bi-level optimal design and power management framework, then presents the formulation of the sizing and power management problem. Section 3 describes the  battery and supercapacitor models. Section 4 gives the details of the devised FLC based on the vectorized fuzzy inference engine. In section 5, the simulation parameters and settings are presented in detail. Section 6 discusses the obtained results, followed by the conclusions in Section 7. 

\section{Bi-Level Optimal Design and Control Framework}\label{sec:BIlevel} 
\begin{figure}[h!tb]
	\centering {\includegraphics[width=\columnwidth]{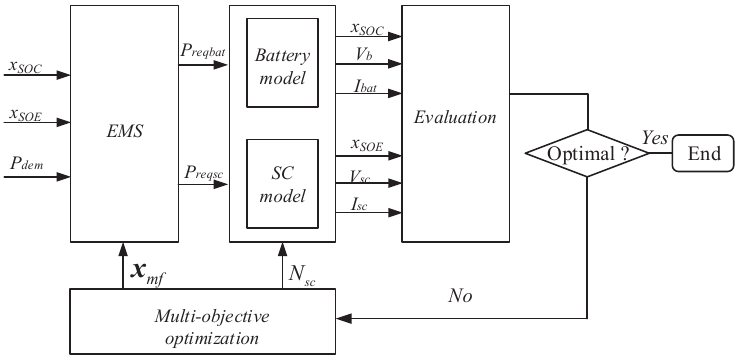}}
	\caption{Framework of the Bi-Level optimal design and control}
	\label{fig:BIconfig}
\end{figure}

The proposed Bi-level optimal design and control framework is presented in Figure \ref{fig:BIconfig}. The term 'Bi-level' here means that the optimization is carried out at both the system integration and controller design levels. The power demand of the driving profile $P_{dem}$, the battery state of charge $ x_{\textit{SOC}} $  and supercapacitor state of energy $ x_{\textit{SOE}} $ are the inputs of the FLC based EMS, while the output of the FLC is the  requested power from the supercapacitor $P_{reqsc}$, and requested power from the battery is calculated by $P_{reqbat} = P_{dem}-P_{reqsc}$. The outputs of the EMS are the inputs of the battery and supercapacitor modeled in Section \ref{sec:batterymodel}, and the evaluation indexes can be calculated based on the outputs of the battery and supercapacitor model. 

The workflow of the Bi-level optimal design and control is illustrated as follows. First, the multi-objective algorithm will generate the sizing parameter matrix and the corresponding static tuning parameter matrix of the power management system. In this work, the  matrices represent the number of supercapacitor banks and the parameters of the  membership functions (MFs) respectively in different pages of the optimization parameters. Second, the FLC-based EMS using the new membership functions  will control the generated  HESS to output the demanded power from the battery and supercapacitor, respectively. Then, the maximum number of laps can be obtained when both the battery and  supercapacitor reach the minimum state of charge values set in the constraints, and the capacity loss of the battery is evaluated based on the average current of the battery during the whole process. Many existing studies have proposed models for capacity loss of lithium-ion batteries. Most existing capacity loss models are validated by discharging the battery at a constant current C rate, and we have not found any experimentally validated models that can predict the battery capacity loss dynamically. Therefore, we choose to estimate the capacity loss of the battery based on the average load like many previous researchers did, and a statistical histogram discharge C rate based approach to evaluate the influences of different C rates on the battery cycle life has also been investigated.  When the Pareto-front of the two evaluation indexes is obtained, the above iteration will terminate; otherwise, it will continue.

The goal of this study is to find the optimal sizing parameter $ {N_{sc}} $  and the parameter vector $ \boldsymbol x_{mf} $ of the membership functions which are  the key parameters of the HESS design and real-time FLC-based EMS,respectively. The optimized EMS will output the requested control command series $ \mathbf u(t)=[P_{reqbat},P_{reqsc}] $ to maximize the number of traveled laps $ J_{laps} $ and battery cycle life $ J_{\textit{lifebat}} $ on a given race circuit:	
\begin{equation}\label{eq:eq1}			
\max J = [J_{laps}(\mathbf x(t), \mathbf u(t), \mathbf p),J_{\textit{lifebat}}(\mathbf x(t), \mathbf u(t),\mathbf p)] 
\end{equation}
\noindent subject to:

\noindent the first-order dynamic constraints
\begin{equation}\label{eq:5optconsns}
{\mathbf{\dot x}}(t) = {\boldsymbol{f}}[{\mathbf{x}}(t),{\mathbf{u}}(t),t,\mathbf {p}],
\end{equation}	
\noindent the boundaries of the state, control and design variables
\begin{equation}\label{eq:boundaries}
\begin{aligned}
\mathbf{x}_{min} \leq {\mathbf{x}(t)}\leq \mathbf{x}_{max}\\
\mathbf{u}_{min} \leq {\mathbf{u}(t)}\leq \mathbf{u}_{max}\\
\mathbf{p}_{min} \leq {\mathbf{p}}\leq \mathbf{p}_{max},
\end{aligned}
\end{equation}

\noindent the algebraic path constraints
\begin{equation}\label{eq:5pathcons}
\boldsymbol{g}_{min} \leq {\boldsymbol{g}}[{\mathbf{x}}(t),{\mathbf{u}}(t),t,\mathbf {p}] \leq \boldsymbol{g}_{max},
\end{equation}	

\noindent and the boundary conditions
\begin{equation}\label{eq:5boncons}
\boldsymbol{b}_{min} \leq \boldsymbol{b} [{\mathbf{x}}({t_0}),{t_0},{\mathbf{x}}({t_f}),{t_f},\mathbf {p}]  \leq  \boldsymbol{b}_{max},
\end{equation}	
where $\boldsymbol{\dot  x} $ is the first-order derivative of the state variables, $ \boldsymbol{f} $ is the dynamic model, $ \mathbf{x} $, $ \mathbf{u} $, $ \mathbf{p}$ are  the state, control and design vector respectively, their lower and upper bounds are $ \mathbf{x}_{min} $, $ \mathbf{u}_{min} $, $ \mathbf{p}_{min} $ and $ \mathbf{x}_{max}, $ $ \mathbf{u}_{max} $, $ \mathbf{p}_{max} $.  While $ \boldsymbol{g} $ and $ \boldsymbol{b} $  are the path and boundary equations respectively with their lower and upper bounds $ \boldsymbol{g}_{min} $, $ \boldsymbol{b}_{min} $  and $ \boldsymbol{g}_{max} $,  $ \boldsymbol{b}_{max} $.

In this work, the algebraic path constraint $ \boldsymbol{g} $ is eliminated by introducing a simple relaxation in Equation (\ref{eq:Nbat}), the state variables $ \boldsymbol x $, control variables $\boldsymbol u $, design parameters $ \boldsymbol p $ and the boundary constraints $ \boldsymbol{b} $ will be presented in the following paragraphs.

\section{Modeling of the HESS}\label{sec:batterymodel} 
\begin{figure}[h!tb]
	\centering {\includegraphics[scale=1.05]{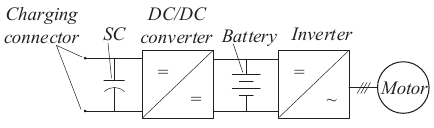}}
	\caption{HESS configuration for an electric race car}
	\label{fig:HESS}
\end{figure}

The HESS configuration for an electric race car is demonstrated in Figure \ref{fig:HESS}.  The supercapacitor (SC) can output and absorb high peak power by controlling a bidirectional DC/DC converter that interfaces the supercapacitor to the DC link of the battery in parallel.  Moreover, the voltage of supercapacitor can be used in a wide range with the help of the DC/DC converter \cite{cao2012new}. There is a  DC/AC inverter between the DC link and the AC motor, which converts direct current  to alternating current. In particular, the DC/AC inverter allows a wide-range input voltage from the DC link.

In this section, the dynamic characteristics of the implemented lithium-ion battery are analyzed first, and a dynamic battery model is employed after comparison. Then, the details of the battery cycle life model are presented. A simplified supercapacitor model is illustrated at the end of this section.  
\subsection{Dynamic Battery Model} 

The most existing battery models for the simulation of battery behavior basically include the semiempirical, electrochemical and electrical ones \cite{Hu2019bbb,Liu2019aaa}. The widely used one is the effective internal resistance (Rint) model, where the voltage and resistance are described as functions of SOC based on experimental data. However, this model with only SOC as independent variable can not capture the influences of the C rate and SOC simultaneously. In order to obtain the optimal sizing parameters and power management strategy for the HESS considering the characteristics of the battery in practical conditions, it is necessary to implement a  proper dynamic battery model that can describe the battery dynamic behavior precisely. In this work, a modified Shepherd model is employed to depict the dynamic characteristics of the battery during the charging and discharging process \cite{Tremblay2007}. The dynamic battery model is represented by  Equation (\ref{eq:eq32}) and Equation (\ref{eq:eq33}) with the assumption that the internal resistance is constant and the thermal behavior of the battery is neglected.  

Discharge:
\begin{equation}\label{eq:eq32}
{V_{bat}} = {E_0} - K\frac{{{Q_{\max }}}}{{{Q_{\max }} - it}}it - K\frac{{{Q_{\max }}}}{{{Q_{\max }} - it}}i - R_{bat}i + A{e^{( - B \cdot it)}}
\end{equation}

Charge:		
\begin{equation}\label{eq:eq33}
{V_{bat}} = {E_0} - K\frac{{{Q_{\max }}}}{{{Q_{\max }} - it}}it - K\frac{{{Q_{\max }}}}{{it - 0.1{Q_{\max }}}}i - R_{bat}i + A{e^{( - B \cdot it)}}
\end{equation}	
\noindent where $ {V_{bat}} $ is the battery voltage ($ V $), $ {E_0} $ is the voltage constant ($ V $), $ K $  is the polarization constant or polarization resistance,  ${Q_{\max }}$ is the total capacity, $ i $  is the battery current, $ R_{bat} $ is the internal resistance. The battery discharge ($i > 0$) or charge ($ i<0 $ ) $ it $  is denoted as		
\begin{equation}\label{eq:eq31}
it = \int i dt.
\end{equation}	

\begin{figure}[h!tb]
	\centerline {\includegraphics[width=\columnwidth]{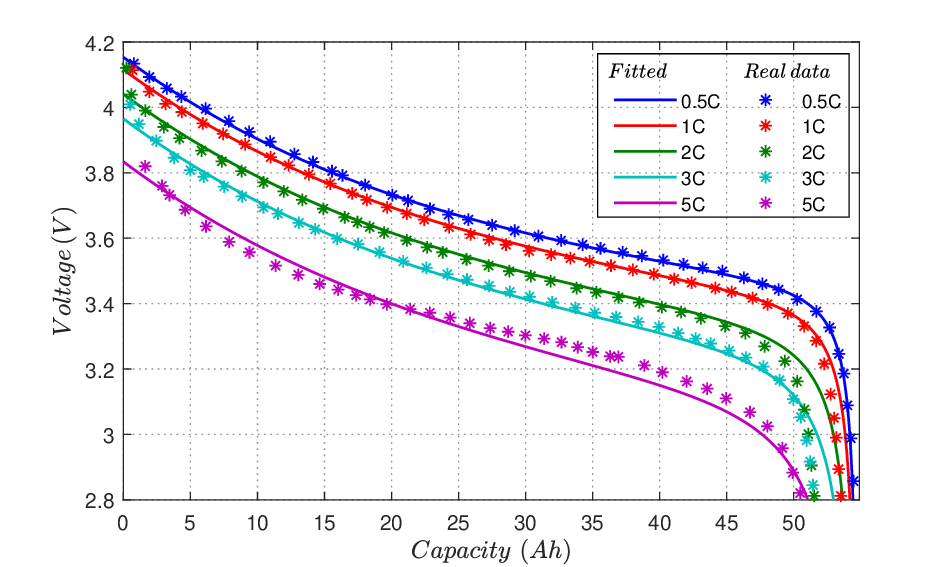}}
	\caption{Comparison of the fitted dynamic battery model and the real data}
	\label{fig:validatebattery}
\end{figure}

The calculation of the voltage amplitude $ A $  ($ V $), time constant inverse $ B $  ($Ah^{-1}$) of the exponential zone, the polarization resistance $ K $  ($ \Omega $) and the voltage  constant $ E_0 $ (V) in Equations (\ref{eq:eq32}) and (\ref{eq:eq33}) are referred to \cite{Tremblay2007}. The  battery model is calibrated  with experimental data, the comparison of the fitted model and the real data is demonstrated as Figure \ref{fig:validatebattery}. The fitted semiempirical model can represent the real battery dynamics satisfactorily.

The state of charge of the battery $ x_{soc} $ and its derivative $\dot x_{soc} $ is denoted as Equation (\ref{eq:xsoc}) and Equation (\ref{eq:dxsoc}) respectively.
\begin{equation}\label{eq:xsoc}
x_{soc} = 100(1-\dfrac{1}{3600Q_{\max}}\int_{0}^{t_{f}}idt)
\end{equation}	
\begin{equation}\label{eq:dxsoc}
\dot x_{soc}= -\dfrac{1}{36Q_{\max}}i
\end{equation}	

The charging/discharging $ i $ is denoted as
\begin{equation}\label{eq:i}
i=\begin{cases}
\dfrac{P_{reqbat}}{N_{bat}V_{bat}\eta_{AD}},&P_{reqbat} \geq 0\\[1em]
\dfrac{P_{reqbat}\eta_{AD}}{N_{bat}V_{bat}},&P_{reqbat} < 0\\
\end{cases}
\end{equation}	

\noindent where $ \eta_{AD} $ is the efficiency of the DC/AC converter taking into account the motor efficiency as a constant value. In this work, the number of  battery cells $N_{bat}$ is determined by the available total mass of the HESS $m_{\textit {HESS}}$ and the number of the supercapacitor banks $N_{sc}$, as shown in Equation (\ref{eq:Nbat}). The total mass of the HESS is fixed in this work given the fact that the mass of a race car is strictly limited in general.
\begin{equation}\label{eq:Nbat}
N_{bat}={\displaystyle \lfloor {(m_{\textit {HESS}}-N_{sc}m_{bank})/m_{cell}}\rfloor}
\end{equation}

\subsection{Battery Cycle Life Model}
In recent years, substantial efforts have been made by both academia and industries to develop models that can predict the degradation of the lithium-ion batteries accurately. Different models were developed to account for various factors responsible for capacity fade such as parasitic side reactions, solid electrolyte interphase formation and resistance increasing. However, experimental data are essential for studying the aging processes of a battery system and verifying the capacity fading mechanisms.  A revised semi-empirical model based on the Arrhenius equation was widely used, and this model was mainly used in optimization and control problems related with batteries \cite{Wang2011}. As shown in Equation (\ref{eq:eq38}) to Equation (\ref{eq:eq41}), the capacity loss of this model is expressed as a function of the discharge current rate $ C_{rate} $, temperature $T$ and ampere-hour throughout ${A_h}$. 
\begin{equation}\label{eq:eq38}
{Q_{loss}} = A_{cl}\exp ({{ - {E_a}} \over {R_{cl}}}){({A_h})^z}
\end{equation}	
\noindent where ${Q_{loss}}$ represents the battery capacity loss, $ A_{cl} $ the pre-exponential factor, ${E_a}$  the activation energy from Arrhenius law ($J$), $ R_{cl} $ is the gas constant of 8.314 , $ T $ is the absolute temperature (K), ${A_h}$ is the Ah-throughput, which represents the amount of charge delivered by the battery during cycling. 

The pre-exponential factor $ A $ in Equation (\ref{eq:eq38}) is proved to be sensitive to the discharge current rate $ C_{rate} $ in the experiments \cite{Wang2011c} , and it is fitted using Equation (\ref{eq:eq39}) in \cite{Shen2014c}.
\begin{equation}\label{eq:eq39}
\ln A_{cl} = a \cdot \exp ( - b \cdot {C_{rate}}) + c
\end{equation}	

The activation energy can be fitted as a linear function of discharge current rate \cite{Wang2011c}, 
\begin{equation}\label{eq:eq40}
{E_a} = d + e \cdot {C_{rate}}
\end{equation}		
\noindent where $ a $, $ b $, $ c $, $ d $, $ e $ are the correction parameters of the battery cycle life model.    

The Ah-throughput can be expressed as 
\begin{equation}\label{eq:eq41}
{A_h} = \int\limits_0^{{t_f}} {{i \over {3600}}dt} 
\end{equation}		

\noindent where $ i $ is the discharge current, $ t_{f} $ is the end time of the current profile.	

In addition to using the average current rate to evaluate  the capacity loss, a statistical method \cite{shen2014} was also implemented to estimate the effect of nonuniform current rate on battery cycle life.

\subsection{Supercapacitor Model}

In this work, the capacity fading of the supercapacitor is neglected because it has much longer cycle life than lithium-ion batteries. The supercapacitor model is simplified to a series connection of a resistance and a supercapacitor bank \cite{ZHANG2018b}. Also, the efficiency of the DC/DC converter between the supercapacitor and the DC link is assumed to be a constant value of 0.95. The recursive supercapacitor model is derived as follows,
\begin{equation}\label{eq:eqSupercapacitor1}
\dot V_{ct} =\begin{cases}
-\dfrac{V_{ct}-\sqrt{V^2_{ct}-4R_{sct}P_{reqsc}/(\eta_{AD}\eta_{dc})}}{2C_{sct}R_{sct}}\,&P_{reqsc} \geq 0\\[1em]
-\dfrac{V_{ct}-\sqrt{V^2_{ct}-4R_{sct}P_{reqsc}\eta_{AD}\eta_{dc}}}{2C_{sct}R_{sct}}\,&P_{reqsc} < 0\\
\end{cases}
\end{equation}
\begin{equation}\label{eq:eqSupercapacitor2}
x_{SOE} = {{V_{ct}}^2 \over {V_{ctmax}^2}}
\end{equation}

\noindent where ${V_{ct}}= {V_c}{N_{sc}}$ is the total open-circuit voltage of the supercapacitor pack assuming that all banks have a uniform behavior, ${t_{k + 1}}$ is the time at step $k+1 $, ${R_{sct}}= {N_{sc}}{R_{sc}}$ is the total equivalent series resistance, ${P_{reqsc}}$ is the demand power from the supercapacitor,  ${\eta _{dc}}$ is the efficiency of the DC/DC converter, ${C_{sct}}={C_{bank}}/{N_{sc}}$ is the total capacity, $x_{SOE}$ is the state of energy, ${V_{ctmax}}$ is the initial open circuit voltage, ${V_c}$ is the open circuit voltage of one supercapacitor, ${N_{sc}}$ is the total number of the banks, ${R_{sc}}$ is the series resistance of one supercapacitor.

The actual total output power of the supercapacitor pack is calculated as
\begin{equation}\label{eq:eqSupercapacitor6}
{P_{sc}} = {V_{ct}} \cdot {{{V_{ct}} - \sqrt {V_{ct}^2- 4{R_{sct}}{P_{reqsc}}/{\eta_{AD} \eta _{dc}}} } \over {2{R_{sct}}}}.
\end{equation}

\section{FLC-based on Vectorized Fuzzy Inference Engine} \label{sec:VIFS}
\begin{figure*}
	\centering {\includegraphics[width=18cm]{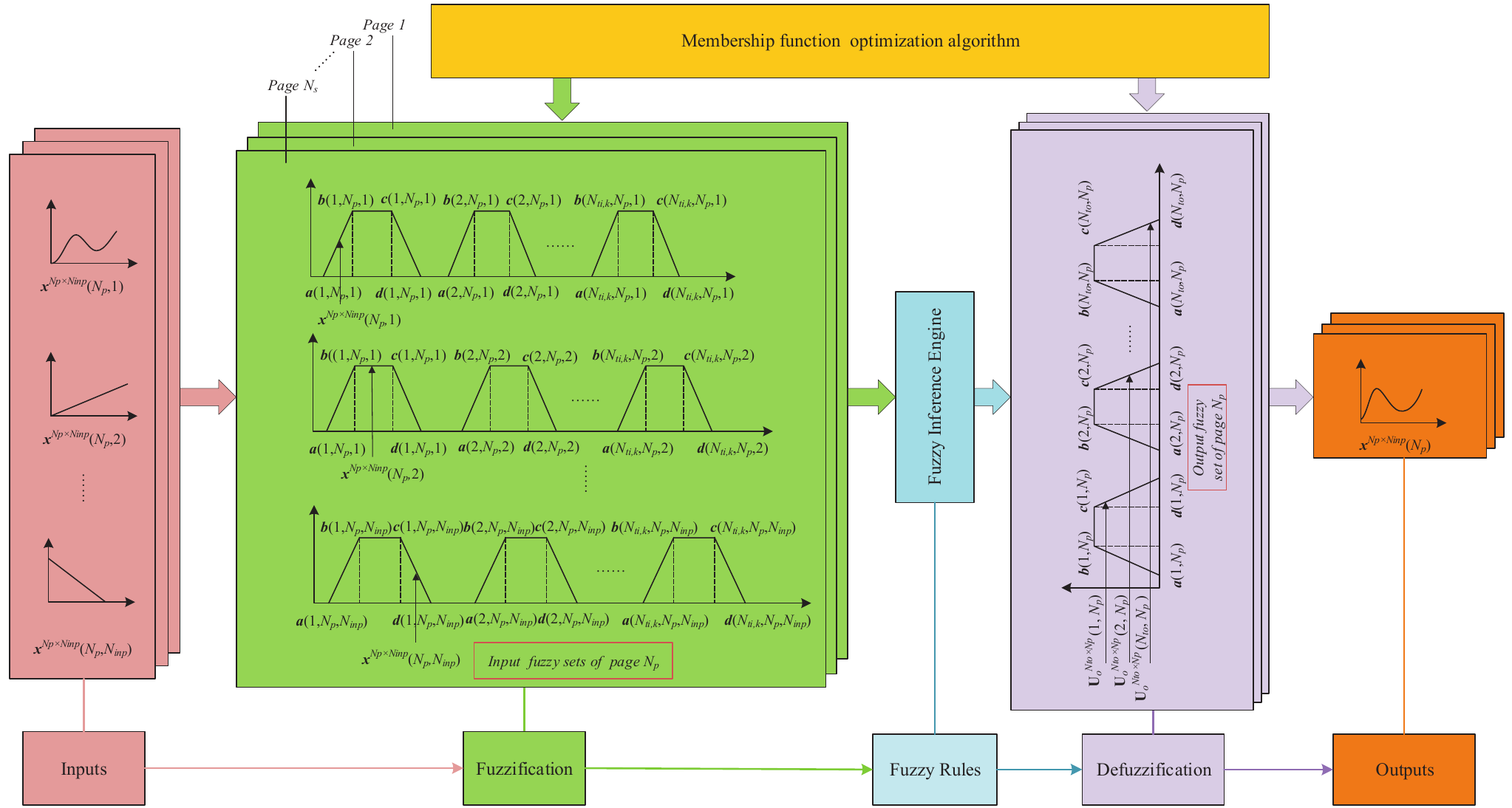}}
	\caption{Framework of the vectorized FLC}
	\label{fig:FLCOPS1}
\end{figure*}

The proposed FLC in this section consists of if-then fuzzy rules, fuzzification, fuzzy inference engine and defuzzification modules. To speed up the optimization and take the advantage of the powerful matrix processing capability of MATLAB, a vectorized fuzzy inference system (VFIS) shown in Figure  \ref{fig:FLCOPS1} is developed for the first time according to the state-of-the-art literature. The developed VFIS is capable of handling  $ {N_{p}} \times {N_{inp}}$ dimensional inputs with $ {N_{p}}$ pages of membership functions each time. This means that $ {N_{p}}$  fuzzy controllers (can be hundreds of thousands depends on the performance of the utilized CPU) can work at the same time with the same page number of inputs and outputs.  
The following paragraph will present the details of fuzzy rules, membership functions, vectorized fuzzification, fuzzy inference engine, and defuzzification operations of the developed vectorized FLC.
\subsection{Fuzzy Rules}
Fuzzy rules are a set of  if-then  linguistic rules used to formulate the conditional relationships that compose a fuzzy logic controller.  For instance, a fuzzy rule can be: if \textit{SOC}  is  \textit{High}  and \textit{SOE}  is  \textit{High}  and  \textit{P\textsubscript{req}} is  \textit{Positive High}  then  \textit{P\textsubscript{sc}} is  \textit{Positive big}. It is reasonable to devise the same if-then rules for the control of different sizes of HESSs since the control objectives of all the HESSs  are the same in this work. The developed fuzzy rules are demonstrated as Figure  \ref{fig:FLC} (a), where the labels N, P, S, M, B means negative, positive, small, medium and big respectively.  The basic idea of the fuzzy rule is to utilize the supercapacitor as a buffer to reduce the high peak power impact on battery and absorb more regenerative braking power.   

\subsection{Membership Functions}

The concept of membership functions was introduced by Zadeh in the first paper on fuzzy sets \cite{ZADEH1965}. A membership function  is a curve or a function that defines how each point of the input variables is mapped to a membership value  between 0 and 1. It is quite challenging to design the optimal MFs for each HESS manually according to the engineering experiences. Besides, considering that the performance of the FLCs are sensitive to their MFs, different MFs of the FLCs with the same fuzzy rules should be devised for different sizes of HESS. Based on these considerations, the parameters of the MFs are selected as parts of the parameters to be optimized in this work.

The trapezoidal-shaped membership function is selected for the fuzzy inference engine based on the considerations that it has high  flexibility \cite{yu2016real}.

\subsection{Vectorized Fuzzification}
During the fuzzification stage, the input variables are identified to the fuzzy sets (membership functions) they belong to and the respective degree of membership  to each relevance will be assigned. For a FIS with  trapezoidal-shaped MFs and a number of $ N_{inp} $ inputs, the fuzzy sets of each can be described using a  matrix $ \boldsymbol{S}_k=[\boldsymbol{a}_k, \boldsymbol{b}_k, \boldsymbol{c}_k, \boldsymbol{d}_k] \in \mathcal{R}^ { {N_p}\times{N_{ti,k}}   \times 4}$. $ N_p $ is the total page number of the inputs; $ N_{ti, k} $ is the number of fuzzy linguistic sets of state input  $ k $, $ k \in \{1,2,...,N_{inp}\}$ and  $a, \ b,\ c,\ d$ are the variables that define one trapezoid. The input matrix is denoted as $ \textbf{x}_{k}\in \mathcal{R}^{N_{p}}$, for $\textbf{X}_{k}$, its membership matrix $ \boldsymbol{\mu}_k \in \mathcal{R}^{{N_{p}} \times N_{ti, k}} $ can be denoted as:

\begin{equation}\label{eq: 4muk}
\left\{ {\begin{aligned}
	&{\boldsymbol{\mu}_k (\boldsymbol{a}_k \leq {\textbf{X}_k} < \boldsymbol{b}_k) = \dfrac{\textbf{X}_k-\boldsymbol{a}_k}{\boldsymbol{b}_k-\boldsymbol{a}_k}} \\ 
	&{\boldsymbol{\mu}_k (\boldsymbol{b}_k \leq {\textbf{X}_k} \leq \boldsymbol{c}_k) = {\boldsymbol{I}}} \\ 
	&{\boldsymbol{\mu}_k (\boldsymbol{c}_k < {\textbf{X}_k} \leq \boldsymbol{d}_k) = \dfrac{\boldsymbol{d}_k-\textbf{X}_k}{\boldsymbol{d}_k-\boldsymbol{c}_k}} 
	\end{aligned}} \right.
\end{equation}

\noindent where $ \boldsymbol{a}_k, \boldsymbol{b}_k, \boldsymbol{c}_k, \boldsymbol{d}_k,  \boldsymbol{I} $ belong to  $ \mathcal{R}^{{N_{p}} \times N_{ti, k}} $, and $ \textbf{X}_k $ is denoted as: 
\begin{equation}\label{eq:4xk}
\textbf{X}_{k}=[\underbrace{\textbf{x}_{k}, \textbf{x}_{k},...,\textbf{x}_{k}}_{N_{ti, k}}] \in \mathcal{R}^{{N_{p}} \times {N_{ti,k}}} 
\end{equation}	

The membership array $ \textbf{U} $  for input $ \textbf{X} $ can be constructed as: 
\begin{equation}\label{eq: 4allU}
\textbf{U}=\{\boldsymbol{\mu}_1,...,\boldsymbol{\mu}_k,..., \boldsymbol{\mu}_{N_{inp}}\} \in   \mathcal{R}^{{N_{p}} \times N_{ti, k}\times N_{inp} }
\end{equation}

\subsection{Vectorized Fuzzy Inference Engine}
Fuzzy inference maps an input space to an output space using fuzzy logic. A FIS tries to formalize the reasoning process of human language by means of fuzzy logic (the built fuzzy If-Then rules). The process of fuzzy inference involves all of the  MFs, If-Then rules, linguistic variables of the inputs and outputs. Mamdani's fuzzy inference method is the most commonly used fuzzy methodology. The search-able fuzzy inference engine is  able to map only one page of the inputs to one page of the outputs. This section will give an elaborate description of the  developed  powerful VFIS  which allows a large number of FLCs operating in parallel based on Mamdani's fuzzy inference method. 	

The linguistic variables are programmed with their integer indexes from the smallest to the biggest in this work. For instance, the fuzzy sets  \{NB, NM, NS, PS, PM, PB\} of the third input  in Figure  \ref{fig:FLC} are  mapped to  $ \{1,2,...,N_{ti,k}\} $ correspondingly, here $ N_{ti,k}=6,\ k=3 $. The fuzzy rule matrix $ \boldsymbol{\Re} \in \mathcal{R}^{N_r\times (N_{inp}+N_{o})}$ is constructed with the mapped integer indexes, $ N_r $  is the number of fuzzy rules and $ N_o $ is the number of outputs. For instance:
\begin{equation}\label{eq:HESSRule}
\begin{array}{l}
{\begin{array}{*{20}{c}}
	{Rule}:\\
	\boldsymbol{\Re}(N_r):\\
	\end{array}{\rm{ }}\begin{array}{*{20}{c}}
	x_{SOC}\\
	1\\
	\end{array}{\rm{ }}\begin{array}{*{20}{c}}
	x_{SOE}\\
	3\\
	\end{array}{\rm{ }}\begin{array}{*{20}{c}}
	P_{req}\\
	6\\
	\end{array}{\rm{ }}\begin{array}{*{20}{c}}
	P_{sc}\\
	6\\
	\end{array}}
\end{array}
\end{equation}	
\noindent where $\boldsymbol{\Re}(N_r)$ denotes the fuzzy rule $ N_r $, it means the rule like: if \textit{SOC}  is  \textit{Low}  and \textit{SOE}  is  \textit{High}  and  \textit{P\textsubscript{req}} is  \textit{Positive High}  then  \textit{P\textsubscript{sc}} is  \textit{Positive High}. The working scheme of the VFIS is illustrated as follows:

\noindent 1) Repeatedly copy the membership matrix $\boldsymbol{\mu}_k  $ into $ N_r $ blocks, and we can obtain:
\begin{equation}\label{eq:HESSmutemp}
\boldsymbol{\mu}_k^{temp}=[\underbrace{\boldsymbol{\mu}_k;\boldsymbol{\mu}_k;...;\boldsymbol{\mu}_k}_{N_r}],\ \boldsymbol{\mu}_k^{temp} \in \mathcal{R}^{N_p\times N_{ti,k}\times N_r}
\end{equation}	

\noindent 2) Create index matrix ${\bf{L}}_{in} \in \mathcal{R}^{N_p\times N_{ti,k}\times N_r}$ for input $ k $: 
\begin{equation}\label{eq:HESSLin}
\resizebox{\columnwidth}{!}{$
\begin{array}{l}
{\bf{L}}_{in} =	\underbrace {\left. {\left\{ {\left[ {\begin{array}{*{20}{c}}
				1\\
				1\\
				\vdots \\
				1
				\end{array}{\rm{ }}\begin{array}{*{20}{c}}
				2\\
				2\\
				\vdots \\
				2
				\end{array}{\rm{ }}\begin{array}{*{20}{c}}
				\cdots \\
				\ldots \\
				\vdots \\
				\cdots 
				\end{array}{\rm{ }}\begin{array}{*{20}{c}}
				{{N_{ti,k}}}\\
				{{N_{ti,k}}}\\
				\vdots \\
				{{N_{ti,k}}}
				\end{array}} \right], \cdots \left[ {\begin{array}{*{20}{c}}
				1\\
				1\\
				\vdots \\
				1
				\end{array}{\rm{ }}\begin{array}{*{20}{c}}
				2\\
				2\\
				\vdots \\
				2
				\end{array}{\rm{ }}\begin{array}{*{20}{c}}
				\cdots \\
				\ldots \\
				\vdots \\
				\cdots 
				\end{array}{\rm{ }}\begin{array}{*{20}{c}}
				{{N_{ti,k}}}\\
				{{N_{ti,k}}}\\
				\vdots \\
				{{N_{ti,k}}}
				\end{array}} \right]} \right\}} \right\}}_{{N_{p}}}{N_{r}}
\end{array}$}
\end{equation}	

\noindent 3) Repeatedly copy the  $ kth $ column of the rule matrix $ \boldsymbol{\Re} \in \mathcal{R}^{N_r\times (N_{inp}+N_{o})}$  into $N_p \times N_{ti,k}$ block arrangement $\boldsymbol{\Re}_{temp} \in  \mathcal{R}^{N_p\times N_{ti,k}\times N_r}$, $k\in\{1,2,\ldots,N_{inp}\}$:
\begin{equation}\label{eq:HESSRtemp}
\boldsymbol{\Re}_{temp}=  \underbrace{\left. {\left[ \begin{array}{cccc}
		\boldsymbol{\Re} _k, &\boldsymbol{\Re} _k, &\ldots, &\boldsymbol{\Re} _k \\
		\boldsymbol{\Re} _k, &\boldsymbol{\Re} _k, &\ldots, &\boldsymbol{\Re} _k \\
		\vdots, &\vdots, &\ldots, &\vdots  \\
		\boldsymbol{\Re} _k, &\boldsymbol{\Re} _k, &\ldots, &\boldsymbol{\Re} _k 
		\end{array} \right]}\right\}}_{{N_{ti,k}}}{N_{p}}
\end{equation}	
\noindent 4) Get the effective membership matrix $\boldsymbol{\mu}_{\textit{eff},k}$ for input $k\in\{1,2,\ldots,N_{inp}\}$:
\begin{equation}\label{eq:HESSmueff}
\boldsymbol{\mu}_{\textit{eff},k}
=\boldsymbol{\mu}_k(L_{in} ==\boldsymbol{\Re}_{temp}),\boldsymbol{\mu}_{\textit{eff},k} \in \mathcal{R}^{N_p\times N_{ti,k}\times N_r}
\end{equation}	

\noindent 5) Combine and get the final membership matrix $\mathbf{U}_{in} \in \mathcal{R}^{N_p\times N_{r}\times N_{inp}}$ for all the input $ \bf{X} $:
\begin{equation}\label{eq:HESSallmu}
\mathbf{U}_{in}=\underbrace{\{\mathop \cup \limits_{j = 1}^{{N_{ti,k}}}\boldsymbol{\mu}_{\textit{eff},k}(j),\mathop \cup \limits_{j = 1}^{{N_{ti,k}}}\boldsymbol{\mu}_{\textit{eff},k}(j),...,\mathop \cup \limits_{j = 1}^{{N_{ti,k}}}\boldsymbol{\mu}_{\textit{eff},k}(j)\}}_{N_{inp}}
\end{equation}	

\noindent 6)  Get the mapped membership matrix $\mathbf{U}_{o}$ for the output fuzzy sets:
\begin{equation}\label{eq:HESSallmuo}
\mathbf{U}_{o}=\mathop \cap \limits_{k = 1}^{{N_{inp}}} \mathbf{U}_{in}(k), \mathbf{U}_o \in \mathcal{R}^{N_p\times N_{r}}
\end{equation}

\noindent 7)  Create index matrix $ L_{o} \in \mathcal{R}^{N_p\times N_r\times N_{to}}$ for output the fuzzy sets:	
\begin{equation}\label{eq:HESSLo}
\resizebox{\columnwidth}{!}{$
\begin{array}{l}
{\bf{L}}_{o} =	\underbrace {\left. {\left\{ {\left[ {\begin{array}{*{20}{c}}
				1\\
				1\\
				\vdots \\
				1
				\end{array}{\rm{ }}\begin{array}{*{20}{c}}
				2\\
				2\\
				\vdots \\
				2
				\end{array}{\rm{ }}\begin{array}{*{20}{c}}
				\cdots \\
				\ldots \\
				\vdots \\
				\cdots 
				\end{array}{\rm{ }}\begin{array}{*{20}{c}}
				{{N_{to}}}\\
				{{N_{to}}}\\
				\vdots \\
				{{N_{to}}}
				\end{array}} \right], \cdots \left[ {\begin{array}{*{20}{c}}
				1\\
				1\\
				\vdots \\
				1
				\end{array}{\rm{ }}\begin{array}{*{20}{c}}
				2\\
				2\\
				\vdots \\
				2
				\end{array}{\rm{ }}\begin{array}{*{20}{c}}
				\cdots \\
				\ldots \\
				\vdots \\
				\cdots 
				\end{array}{\rm{ }}\begin{array}{*{20}{c}}
				{{N_{to}}}\\
				{{N_{to}}}\\
				\vdots \\
				{{N_{to}}}
				\end{array}} \right]} \right\}} \right\}}_{{N_{p}}}{N_{r}}
\end{array}$}
\end{equation}			

\noindent 8) Repeatedly copy the column of output fuzzy sets in the rule matrix $ \boldsymbol{\Re} \in \mathcal{R}^{N_r\times (N_{inp}+N_{o})}$  into a $N_p \times N_{to}$ block arrangement $\boldsymbol{\Re}_{otemp} \in  \mathcal{R}^{N_p\times N_r\times N_{to}}$, $N_{to}$ is the number  of fuzzy linguistic sets of output :
\begin{equation}\label{eq:HESSRotemp}
\boldsymbol{\Re}_{otemp}=  \underbrace{\left. {\left[ \begin{array}{cccc}
		\boldsymbol{\Re} _{N_{o}}, &\boldsymbol{\Re} _{N_{o}}, &\ldots, &\boldsymbol{\Re} _{N_{o}} \\
		\boldsymbol{\Re} _{N_{o}}, &\boldsymbol{\Re} _{N_{o}}, &\ldots, &\boldsymbol{\Re} _{N_{o}} \\
		\vdots, & \vdots, &\ldots, &\vdots  \\
		\boldsymbol{\Re} _{N_{o}},&\boldsymbol{\Re} _{N_{o}}, &\ldots, &\boldsymbol{\Re} _{N_{o}}
		\end{array} \right]}\right\}}_{{N_{to}}}{N_{p}} 
\end{equation}	

\noindent 9) Repeatedly copy the membership matrix of the output fuzzy sets $ \mathbf{U}_o \in \mathcal{R}^{N_p\times N_{r}} $ into $ N_{to}$ blocks $\mathbf{U}_{o,temp}$:
\begin{equation}\label{eq:HESSmuoutrap}
\mathbf{U}_{o,temp}=\underbrace{\{\mathbf{U}_o,\mathbf{U}_o,..., \mathbf{U}_o\}}_{N_{to}},\mathbf{U}_{o,temp} \in \mathcal{R}^{N_p\times N_{r}\times N_{to}}
\end{equation}

\noindent 10) Get the effective membership matrix $\mathbf{U}_{\textit{eff},o}$ of all the output fuzzy sets:
\begin{equation}\label{eq:HESSmuout}
\mathbf{U}_{\textit{eff},o}=\mathbf{U}_{o,temp}(L_{o} ==\boldsymbol{\Re}_{otemp}),\mathbf{U}_{\textit{eff},o} \in \mathcal{R}^{N_p\times N_{r}\times N_{to}}
\end{equation}	

\noindent 11) Merge the membership matrix of the output fuzzy sets in all the fuzzy rules
\begin{equation}\label{eq:HESSMuO}
\mathbf{U}_{o,\textit{final}}=\bigcup\limits_{i = 1}^{N_{r}}  \mathbf{U}_{o,\textit{eff}}(i), \mathbf{U}_{o,\textit{final}} \in \mathcal{R}^{N_p\times N_{to}}
\end{equation}		

By the above calculation, the membership of each trapezoid of the output fuzzy set is obtained as $\mathbf{U}_{o,\textit {final}}$, and the next step is the defuzzification.

\subsection{Vectorized Defuzzification}
\begin{figure}[h!tb]
	\centering {\includegraphics[width=\columnwidth]{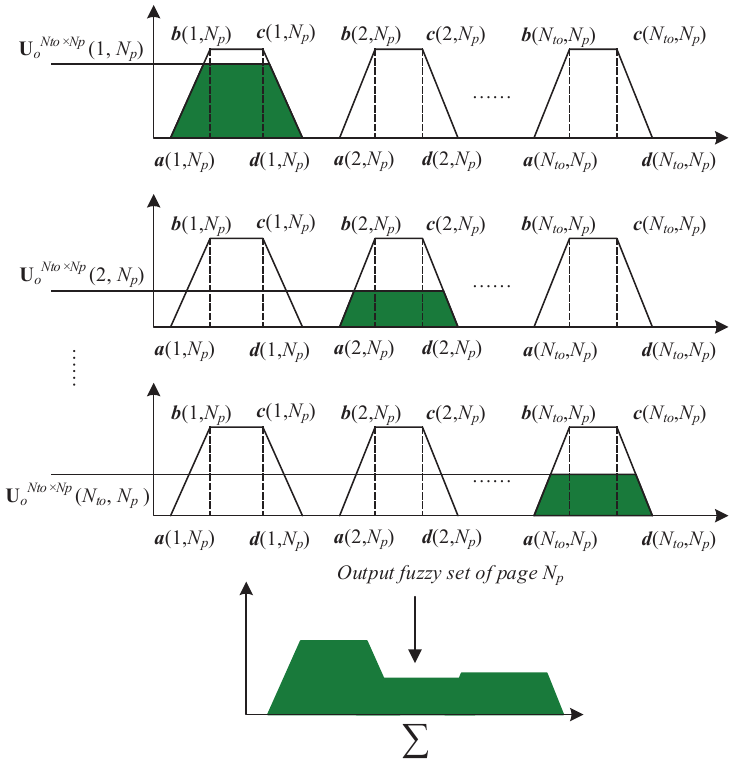}}
	\caption{Defuzzification  process based on the center of gravity method}
	\label{fig:FLCOPS3}
\end{figure}

The purpose of the defuzzification process is to produce a quantifiable result in crisp logic based on the given fuzzy sets and corresponding membership degrees. The defuzzification  process based on the center of gravity method is demonstrated in Figure \ref{fig:FLCOPS3}. 

The procedure is elaborated as followings:
\noindent 1) Discrete the output fuzzy sets into $ N_{dis} $  parts  $ \boldsymbol{x}_o \in \mathcal{R}^{N_{dis}} $ from its minimum value ${\boldsymbol{x}_{o,\min }}$ to the maximum one ${\boldsymbol{x}_{o,\max }}$,
\begin{equation}\label{eq:HESSxo}
\boldsymbol{x}_o = [{\boldsymbol{x}_{o,\min}}:{{({\boldsymbol{x}_{o,\max }} - {\boldsymbol{x}_{o,\min }})} \mathord{\left/
		{\vphantom {{({\boldsymbol{x}_{o,\max }} - {\boldsymbol{x}_{o,\min }})} {({N_{dis}} - 1):{\boldsymbol{x}_{o,\max }}}}} \right.
		\kern-\nulldelimiterspace} {({N_{dis}} - 1):{\boldsymbol{x}_{o,\max }}}}]
\end{equation}	
\noindent where the value of $N_{dis}$ affects the accuracy of the crisp output, for instance, the increasing of $N_{dis}$ will improve the precision but will increase the computational burden.
\\

\noindent 2) Repeatedly copy $\boldsymbol{x}_o \in \mathcal{R}^{N_{dis}} $  and output fuzzy set $ \boldsymbol{S}_o=[\boldsymbol{a}_o, \boldsymbol{b}_o, \boldsymbol{c}_o, \boldsymbol{d}_o] \in \mathcal{R}^ {{N_{to}} \times {N_p}\times 4}$, we can obtain $\boldsymbol{x}_{o,temp} \in \mathcal{R}^ {{N_{to}} \times {N_p}\times {N_{dis}}} $ and $\boldsymbol{S}_{o,temp} \in \mathcal{R}^ {{N_{to}} \times {N_p}\times {N_{dis}}} $ respectively:
\begin{equation}\label{eq:HESSxotemp}
\boldsymbol{x}_{o,temp}=  \underbrace{\left. {\left[ \begin{array}{cccc}
		\boldsymbol{x} _o, &\boldsymbol{x} _o, &\ldots, &\boldsymbol{x} _o \\
		\boldsymbol{x} _o, &\boldsymbol{x} _o, &\ldots, &\boldsymbol{x} _o \\
		\vdots, & \vdots, &\ldots, &\vdots  \\
		\boldsymbol{x} _o,&\boldsymbol{x} _o, &\ldots, &\boldsymbol{x} _o
		\end{array} \right]}\right\}}_{{N_{to}}}{N_{p}}
\end{equation}	

\begin{equation}\label{eq:HESSSotemp}
\boldsymbol{S}_{o,temp}= \underbrace{[\boldsymbol{S}_o,\boldsymbol{S}_o,...,\boldsymbol{S}_o]}_{N_{dis}}
\end{equation}

\noindent 3) Calculate the membership matrix of  $ \boldsymbol{x}_{o,temp} $ based on the output fuzzy set $ \boldsymbol{S}_{o,temp}=[\boldsymbol{a}_{o,temp}, \boldsymbol{b}_{o,temp}, \boldsymbol{c}_{o,temp}, \boldsymbol{d}_{o,temp}] \in \mathcal{R}^ {{N_{to}} \times {N_p} \times{N_{dis}}\times 4}$:	
\begin{equation}\label{eq: 4muko}
\left\{ {\begin{aligned}
	&{\boldsymbol{\mu}_{o} (\boldsymbol{a}_{o,temp} \leq {\boldsymbol{x}_{o,temp}} < \boldsymbol{b}_{o,temp}) = \dfrac{\boldsymbol{x}_{o,temp}-\boldsymbol{a}_{o,temp}}{\boldsymbol{b}_{o,temp}-\boldsymbol{a}_{o,temp}}} \\ 
	&{\boldsymbol{\mu}_{o,temp} (\boldsymbol{b}_{o,temp} \leq {\boldsymbol{x}_{o,temp}} \leq \boldsymbol{c}_{o,temp}) = {\boldsymbol{I}}} \\ 
	&{\boldsymbol{\mu}_o (\boldsymbol{c}_{o,temp} < {\boldsymbol{x}_{o,temp}} \leq \boldsymbol{d}_{o,temp}) = \dfrac{\boldsymbol{d}_{o,temp}-\boldsymbol{x}_{o,temp}}{\boldsymbol{d}_{o,temp}-\boldsymbol{c}_{o,temp}}} 
	\end{aligned}} \right.
\end{equation}	

\noindent 4) Repeatedly copy the membership matrix $\mathbf{U}_{o,final} $ in to $ N_{dis} $ blocks $\mathbf{U}_{o,temp} \in \mathcal{R}^ {{N_{to}} \times {N_p}\times {N_{dis}}} $:
\begin{equation}\label{eq:HESSuocopy}
\mathbf{U}_{o,temp}=\underbrace{[\mathbf{U}_{o,final},\mathbf{U}_{o,final},...,\mathbf{U}_{o,final}]}_{N_{dis}}
\end{equation}	

\noindent 5) Find the effective membership matrix:
\begin{equation}\label{eq:HESSuoeff}
\mathbf{U}_{\textit{eff},o}=  \mathbf{U}_{o,temp}\bigcap \boldsymbol{\mu}_{o}, \mathbf{U}_{\textit{eff},o} \in \mathcal{R}^ {{N_{to}} \times {N_p}\times {N_{dis}}} 
\end{equation}	

\noindent 6) Merge the  membership matrix obtained in last step:
\begin{equation}\label{eq:HESSUoFinal}
\mathbf{U}_{o,x}=\bigcup\limits_{i = 1}^{N_{r}}	\mathbf{U}_{\textit{eff},o}(i), \mathbf{U}_{o,x} \in \mathcal{R}^ {{N_p}\times {N_{dis}}} 
\end{equation}


\noindent 7) Calculate the crisp output matrix for all the input matrices:
\begin{equation}\label{eq:HESSy}
y=\dfrac{\sum\limits_{i = 1}^{{N_{dis}}}\boldsymbol{x}_o(i) \circ\mathbf{U}_{o,x}\left({\boldsymbol{x}_o}(i)\right)}{\sum\limits_{i = 1}^{{N_{dis}}}{\mathbf{U}_{o,x} \left({\boldsymbol{x}_o}(i)\right)}}, y\in \mathcal{R}^ {N_p}
\end{equation}

In order to design the fuzzy rules and membership functions conveniently, the devised vectorized FLC modules  illustrated above are developed in  MATLAB with standard and user-friendly interfaces. 

\section{Simulation parameters and Settings}\label{sec:cons} 
\begin{figure}[h!tb]
	\centering {\includegraphics[width=\columnwidth]{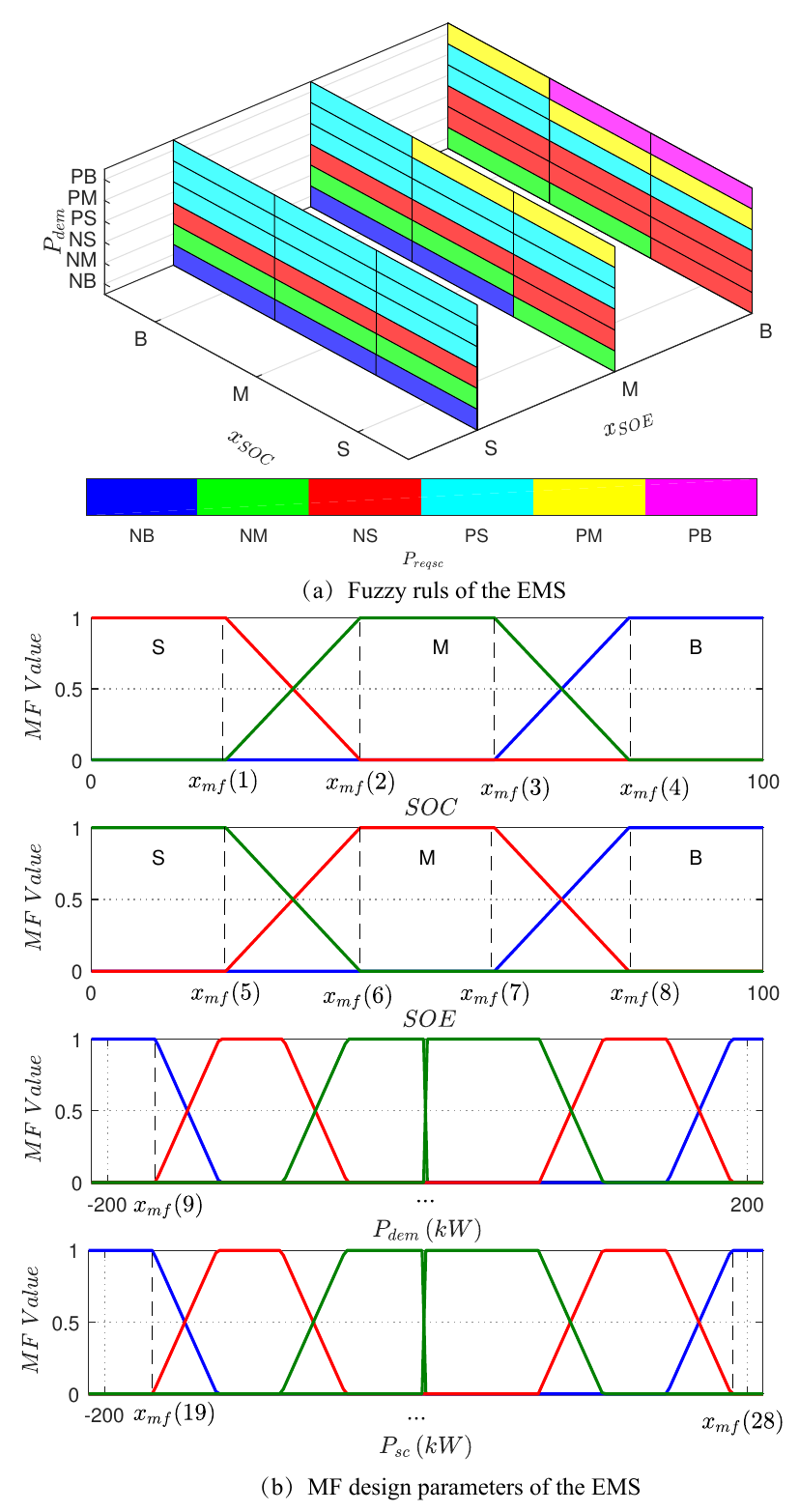}}
	\caption{Parameters of the FLC based EMS}
	\label{fig:FLC}
\end{figure}

The state variables include the battery state of charge $x_{\textit{SOC}}$ and state of  energy of the supercapacitor $x_{\textit{SOE}}$, $ \mathbf{x}=[x_{\textit{SOC}},\ x_{\textit{SOE}}] $. The control variable output by the FLC in this work is the requested power from the supercapacitor $ \mathbf{u}=P_{reqsc}$, while the demand power from the battery can be calculated by $P_{reqbat}=P_{dem}-P_{reqsc}$. The design parameter vector is $ \mathbf{p} =\{ N_{sc}, \boldsymbol{x}_{mf}\}$. The designed fuzzy rules and initial membership functions are demonstrated in Figure \ref{fig:FLC}, and there are 28 parameters of the devised membership functions plus one design parameter of the HESS in one page of parameters to be optimized. The design vector $\mathbf{p}$ is constrained by defining $\mathbf{p}_{min}$ and $\mathbf{p}_{max}$.

\begin{figure}[h!tb]
	\centering {\includegraphics[width=\columnwidth]{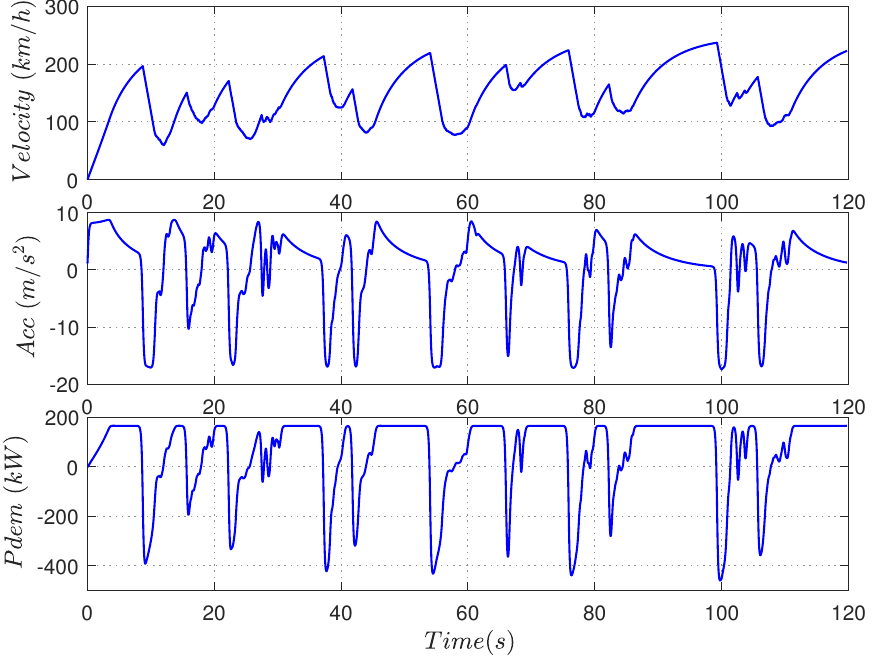}}
	\caption{Power demand in Nurburgring circuit}
	\label{fig:DemandPower}
\end{figure}

The operating profile of an electric race car is quite different from the one of a conventional electric vehicle running on a city road. Thus, standard driving cycles are not suitable for research on electric race cars. The real driving cycle of a race car in the Nurburgring circuit is chosen as the test scenario.  The demand driving/braking power is calculated by Equation (\ref{eq:4}). The corresponding velocity profile, acceleration profile and demand power are demonstrated as Figure \ref{fig:DemandPower}.
\begin{equation}\label{eq:4}
P_{dem} = (\frac{1}{2}\rho {C_d}A{v^2} + fm_vg + m_va)v
\end{equation}	

The detailed simulation parameters of the race car, 53 Ah (Rated) high energy lithium-ion battery, 2.85V/3400F high-performance supercapacitor and the converters are illustrated in Table \ref{tab:ParameterBatteryV}. 
\begin{table}[h!tb] 
	\centering	
	\caption{Parameter values of the simulation}	\label{tab:ParameterBatteryV}
	\centering 
	\vskip 0.2cm
	\resizebox{\columnwidth}{!}{
		\begin{tabular}{c c c}
			\hline	
			\hline	
			\textbf{\textit{Parameters}}&\textbf{\textit{Symbol}}& \textbf{\textit{Value}}\\
			\hline
			Vehicle mass (kg)& $ m_v $  & 570\\			
			Aerodynamics coefficient  ($h^2N/ {km}^2 $) &$ \rho C_dA $ &0.075\\			
			Rolling resistance coefficient  &$ f $  &	0.016\\		
			Mass of the battery cell ($ kg $)		& $ m_{cell} $ & 1.15\\	
			Voltage constant of the battery cell ($ V $)		& $ E_0 $ & 3.43\\		
			Maximum capacity of the battery cell(Ah)& $ Q_{max} $ &55\\		
			Polarization resistance of the battery cell  ($ \Omega $)&$ K $ &8.85$ \times 10^{-5}$  \\
			Internal resistance of the battery cell ($ \Omega $)& $ R $ &1.33$ \times 10^{-3}$  \\
			Voltage amplitude of the battery cell ($ V $ )&A&0.761\\
			Time constant inverse of the battery cell ($Ah^{-1}$)&$ B $ & 0.040\\
			Fitting parameter of the pre-exponential factor&a&1.345\\
			Fitting parameter of the pre-exponential factor&b&0.2563\\
			Fitting parameter of the pre-exponential factor&c&9.179\\
			Fitting parameter of activation energy&d&46868\\
			Fitting parameter of activation energy&e&-470.3\\	
			Mass of the supercapacitor bank ($ kg $)		& $ m_{bank} $ & 0.52\\			
			Supercapacitor 	bank capacity($F$) &$ C_{bank} $ &3400\\   
			Supercapacitor equivalent series resistance ($\Omega $)&$ R_s $ &2.2$ \times 10^{-4}$  \\
			DC/DC converter efficiency&$\eta_{dc}$&0.95\\	
			DC/AC converter efficiency&$\eta_{AD}$&0.96\\			    		    		    		    		    			    		    		    		    		    
			\hline
			\hline
		\end{tabular}
	}	
\end{table}

In the FLC based EMS, the SOE and current of the supercapacitor are constrained between 0.1 and 0.99, -2000 A and 2000A respectively. While the SOC  of the lithium-ion battery is constrained between 0.2 and 0.9, the current is regulated by adjusting the requested power from the battery. When the lithium-ion battery is exhausted, the simulation of one iteration will be terminated and the objective functions will be  evaluated.  The temperature is for sure very important in any kind of vehicle equipped with batteries since it can affect the performance of the batteries directly. However, it is very difficult to model the heat generation, dispassion and the thermal control system of the energy storage system on an electric vehicle precisely. Actually, it is reasonable to assume that the temperature is controlled at a constant value (23 $ ^oC $) by adjusting the thermal control system \cite{Hu2015,Ebbesen2012}.

In this work, a controlled elitist NSGA which is a variant of NSGA-II \cite{Deb2011a} is implemented  to solve the  multi-objective optimization problem. Instead of only choosing the top-ranking non-dominated fronts, the controlled elitist GA also favors individuals that can assist to improve the diversity of the population even if their fitness values are relatively lower.
\section{Results}

\begin{figure}[h!tb]
	\centering {\includegraphics[width=\columnwidth]{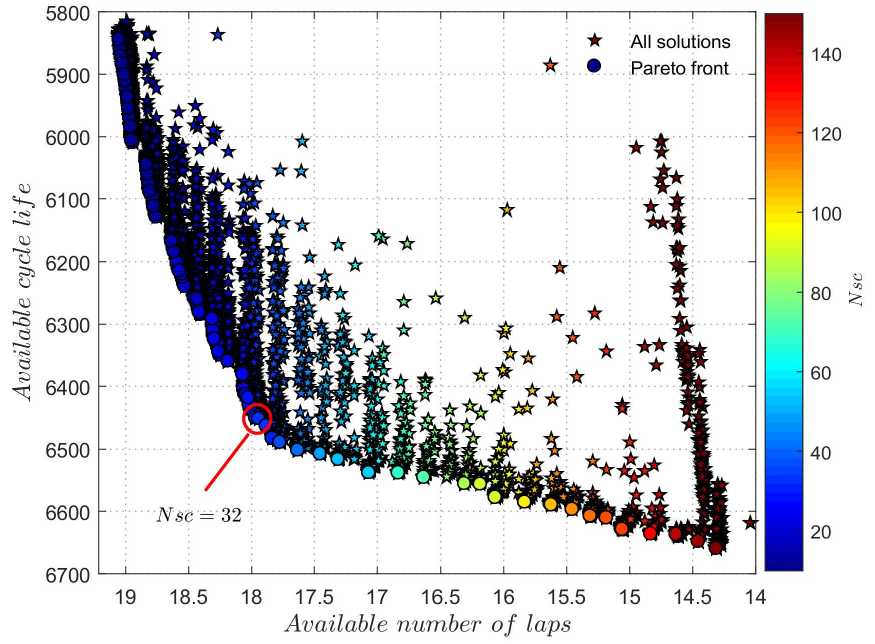}}
	\caption{Multi-objective sizing and control solutions when $ m_{\textit{HESS}} =320 \, kg$ }
	\label{fig:Pareto}
\end{figure}

In this section, the results of the multi-objective optimal sizing and control of the HESS are presented and analyzed in detail.  Figure \ref{fig:Pareto} presented the achieved results when the total mass of the HESS is limited to 320 $ kg $ and the average current is used to evaluate the capacity loss. The population size is set to 500 in the NSGA-II optimization algorithm, and the optimization is terminated after about 13 hours in a ThinkPad T470P laptop with Intel(R) Core(TM) i5-7300HQ CPU @2.50 GHz CPU and 16GB RAM.  The  number of total iterations is 1839, which means that about 0.92 million solutions have been evaluated during the optimization process. All of the solutions demonstrated in Figure \ref{fig:Pareto} are associated  with corresponding design and control parameters.

From the sizing point of view, using a different number of supercapacitors means different compromises between high power density and high energy density. As it is demonstrated in Figure \ref{fig:Pareto},  utilizing more supercapacitors can assist to reduce the average current of the lithium-ion battery which is beneficial for longer cycle life  of the battery, but reduce the energy density of the HESS which results in shorter driving mileage. When fewer supercapacitors are used, the results will be the opposite. It is also observed in Figure \ref{fig:Pareto}  that HESS with the same design solutions (makers filled with the same color) may achieve different values of both objective functions, which means that for the same HESS with uniform fuzzy rules, the parameters of the membership functions will determine whether we can achieve the Pareto optimal solutions. Thanks to the proposed Bi-level optimal sizing and control framework, the corresponding sizing parameter $N_{sc}$ of each HESS and the membership function parameters $ \mathbf x_{mf}$  of the related EMS are coupled and obtained at the same time for all the solutions  including those on the Pareto front. 

\begin{figure}[h!tb]
	\centerline {\includegraphics[width=\columnwidth]{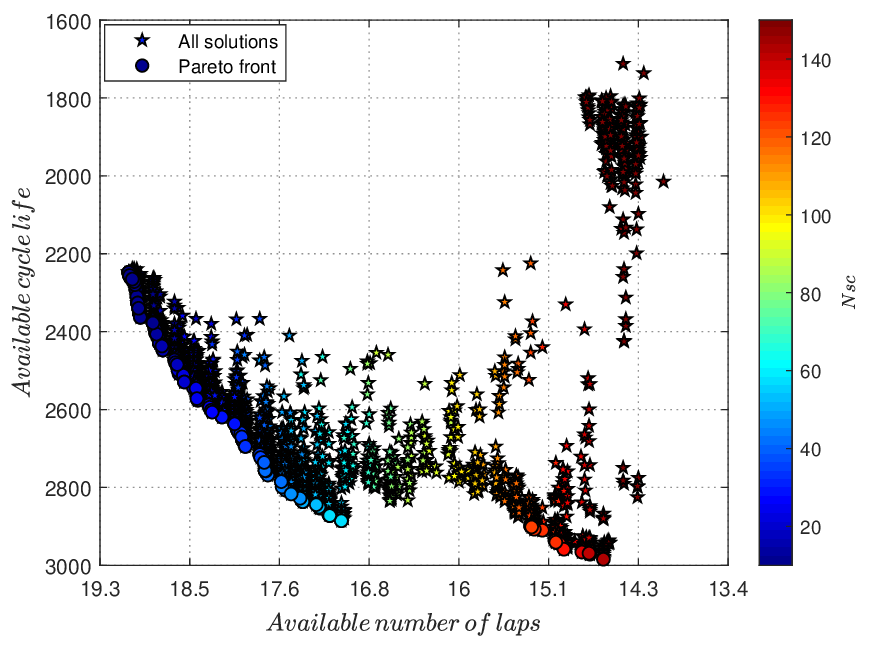}}
	\caption{Pareto solutions using nonuniform C rate when $ m_{\textit{HESS}} =320 \, kg$}
	\label{fig:ParetoViaCrate}
\end{figure}

\begin{figure}[h!tb]
	\centerline {\includegraphics[width=\columnwidth]{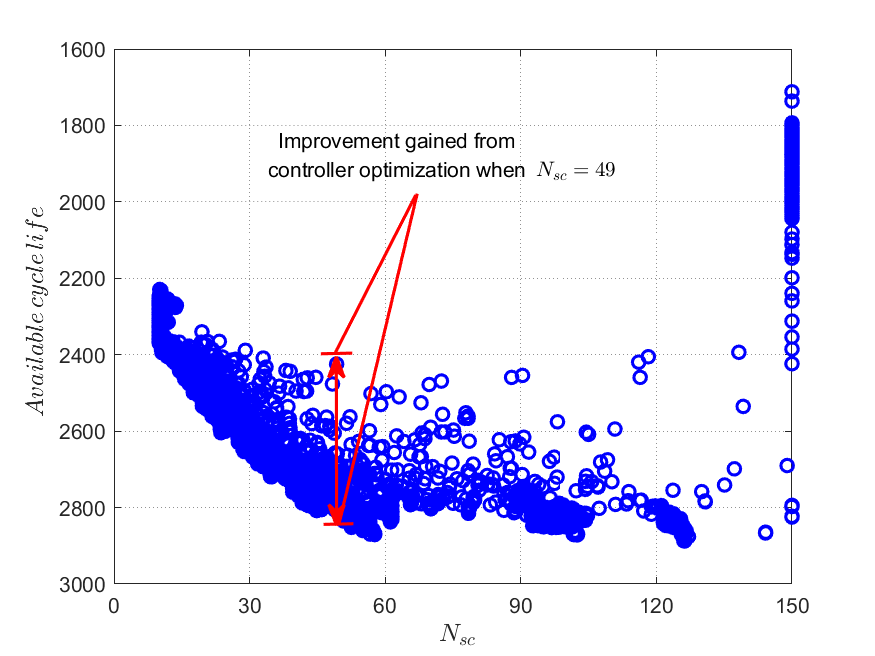}}
	\caption{ Battery cycle life improvement by optimizing the FLC based EMS}
	\label{fig:NscvsCyclelife}
\end{figure}

From Figure \ref{fig:ParetoViaCrate}, we can see that, when utilize the statistical method to estimate the effect of nonuniform current rate on battery cycle life, the battery cycle life is between 1717 and 2984. In particular, the HESS with 50 supercapacitors can improve the battery cycle life by $27.5 \%$.  The improvements of the available battery cycle life by optimizing the membership functions of the FLC based EMS have been further demonstrated in detail in Figure \ref{fig:NscvsCyclelife}. The gained improvement of the battery cycle life from controller optimization when $N_{sc}=46$ is $15.1 \%$.

\begin{figure}[h!tb]
	\centering {\includegraphics[width=\columnwidth]{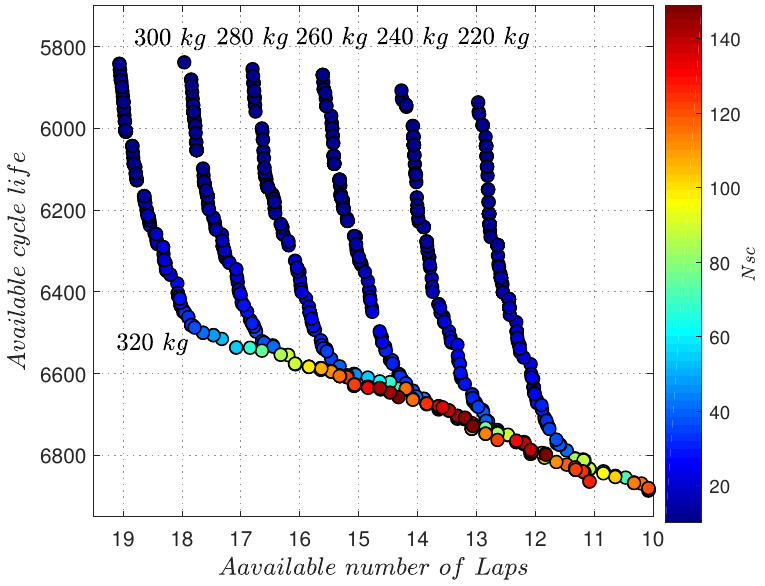}}
	\caption{Pareto optimal solutions for HESS with different  mass $m_{\textit{HESS}}$}
	\label{fig:MultiPareto}
\end{figure}

Moreover, this work has investigated the optimal sizing and control results of HESSs with different total mass. From Figure \ref{fig:MultiPareto}, we can draw the following basic conclusions: 1) HESSs with smaller total mass will cover a fewer number of available laps, but the available cycle life of the battery is longer due to their shorter operating mileage; 2) We can achieve a pretty decent compromised solution that can enhance both  objective functions with only about 40 supercapacitor banks and the optimized membership functions.  

\begin{figure}[h!tb]
	\centering {\includegraphics[width=\columnwidth]{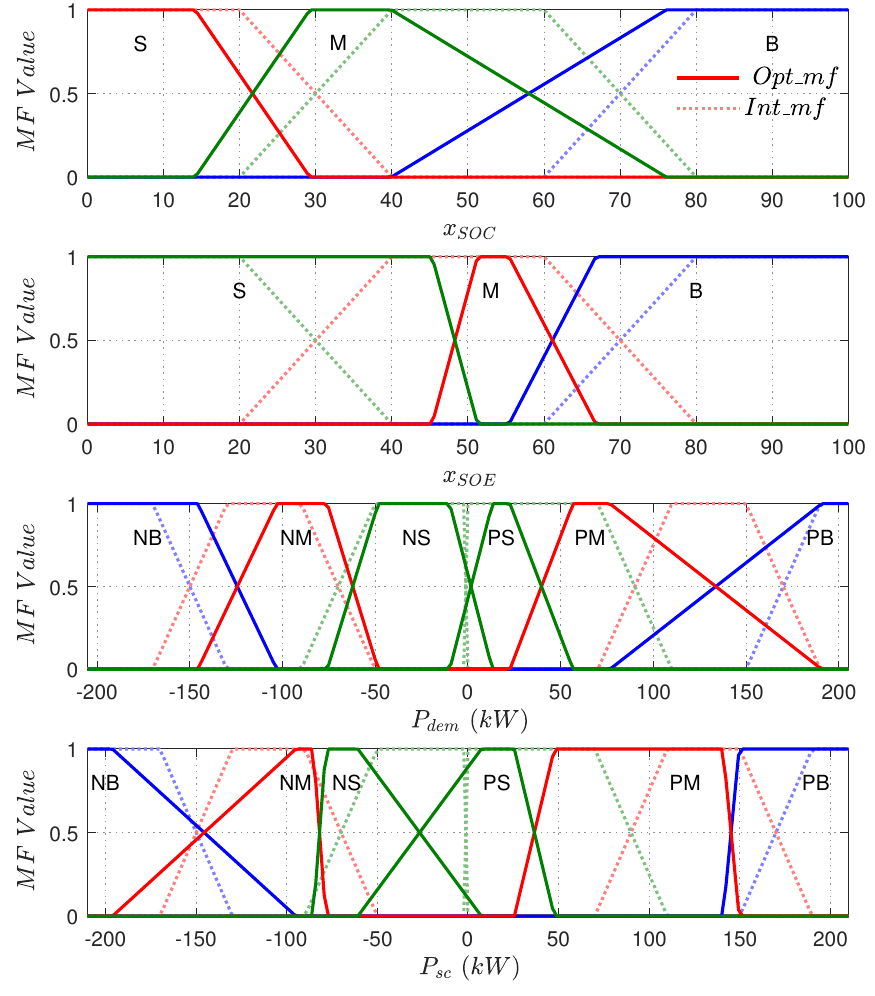}}
	\caption{The initial and optimized membership functions when $N_{sc}=32$}
	\label{fig:MFcomp}
\end{figure}

In order to analyze the reason for the exhibited advantages of the proposed Bi-level optimal sizing and control framework, one solution from the Pareto front in Figure \ref{fig:Pareto} ($N_{sc}=32$) is compared with the solution with the same sizing parameter but the initially devised membership functions. Figure \ref{fig:MFcomp} demonstrates the initial and optimized membership functions with the dotted lines and solid lines respectively. 

\begin{figure}[h!tb]
	\centering {\includegraphics[width=\columnwidth]{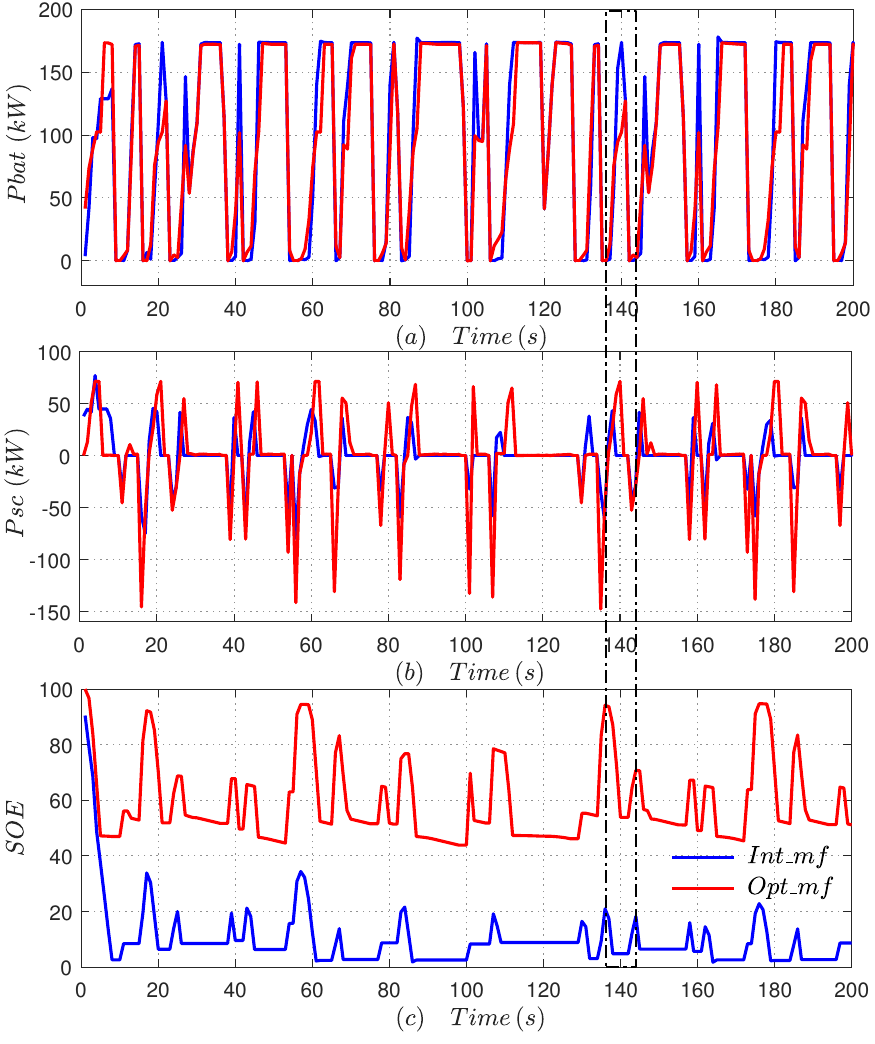}}
	\caption{Pareto optimal solutions for HESS with different MFs}
	\label{fig:detailcomp}
\end{figure}

The achieved available number of laps of the initial and optimized solutions are very similar which are respectively 17.88 and 17.98. This is mostly due to the fact that the two cases are implemented with the same HESS and the available mileage is mainly determined by the sizing parameters rather than the control parameters. However, the  available cycle life of the battery is different which are respectively 6082 and 6463. This means that HESS with the optimized membership functions improved the battery cycle life by 6.3\%. Figure \ref{fig:detailcomp} presents the interested variables between 0-200s, as it is illustrated in Figure \ref{fig:detailcomp} (a) and Figure \ref{fig:detailcomp} (b), the EMS with initially devised membership functions tends to request more high peak power from the battery and less from the supercapacitors which will accelerate the degradation  of the battery. This phenomenon can be explained with the curve of  \textit{SOE}    in Figure \ref{fig:detailcomp} (c). We can see that EMS with the initial devised membership functions tends to exhaust the supercapacitors very fast at a few seconds after starting the operation and the average  \textit{SOE}  is under 20\% during the simulation which is not capable to provide long-time high peak power to protect the battery. While EMS with the optimized membership functions tends to maintain the  \textit{SOE}  of the supercapacitors above 50\%, which helps to play the role of shaving the peak and filling the valley very well during the whole driving profile. For instance, the curves in the dotted box in Figure \ref{fig:detailcomp} (c) demonstrate that the requested power from the battery is less  after optimizing the MFs since the \textit{SOE} is maintained at a relatively high level due to the optimized EMS. 



\section{Conclusions}
More supercapacitors do not always guarantee a better overall  performance especially  when the total mass of the hybrid energy storage system is limited due to the fact that the energy density of the supercapacitor is quite poor and it can be exhausted very fast even if it has high power density. However, we are able to obtain a pretty good balanced performance with less supercapatitors and the optimized power management system by the proposed optimization framework. The proposed Bi-level optimal sizing and control framework in this work makes it possible to obtain the global optimal solutions since it enables the optimization algorithm to search both the design and control parameters simultaneously. The user could choose the favored sizing solution from the obtained Pareto front packaged with the optimal membership functions based on a preferred compromise between the two objectives.  The obtained global optimal sizing parameters and optimal parameters of the real-time controller on the Pareto front can be put into real-time implementations. In addition to the Bi-level optimal sizing and control framework, the devised vectorized fuzzy inference system with standard interfaces can be used in other kinds of real time feedback control problems, in particular, it can assist to dramatically improve the computational efficiency when needs to optimize the parameters of  fuzzy logic controller. 

\ifCLASSOPTIONcaptionsoff
  \newpage
\fi

\end{document}